\documentclass[twocolumn,aps,pra]{revtex4}
\usepackage[dvips]{graphics,graphicx}
\usepackage{amsmath}
\begin{document}

\title{NOON state of Bose atoms in the double-well potential via an excited state quantum phase transition}
\author{A. A. Bychek$^{1,2}$}
\author{D. N. Maksimov$^{1,3}$}
\author{A. R. Kolovsky$^{1,2}$}
\affiliation{$^1$Kirensky Institute of Physics, Federal Research Center KSC SB RAS, 660036, Krasnoyarsk, Russia\\
$^2$Siberian Federal University,  660041, Krasnoyarsk, Russia \\
$^3$Reshetnev Siberian State University of Science and Technology, 660037, Krasnoyarsk, Russia}
\date{\today}
\begin{abstract}
We suggest a simple scheme for creating a NOON state of repulsively interacting Bose atoms in the double-well potential. The protocol consists of two steps. First, by setting atom-atom interactions to zero, the system is driven to the upper excited state. Second, the interactions is slowly increased and, simultaneously,  the inter-well tunneling is decreased to zero. We analyze fidelity of the final state to the NOON state depending on the number of atoms, ramp rate, and fluctuations of the system parameters. It is shown that for a given fidelity the ramp rate scales algebraically with the number of atoms.
\end{abstract}
\maketitle

\section{Introduction}
\label{sec0}

Non-classical states of bosonic ensembles play important role in
quantum computing, measurement, and communication
\cite{Wine13,Lee12,Pezz16}. Among many different implementations
\cite{Monr96,Halj05,McDo07,Lo15,Fisc15,Kien16,Zhan16} the two-mode
Bose-Hubbard model
\cite{Smer97,Cira98,Spek99,Java99,Meno01,Hine03,Ho04,Albi05,Pezz05,Lee06,Huan06,Muel06,Gati07,Stre07,Teic07,Este08,Khod09,Haig10,Mazz11,Garc12,Dell12,Java14,Volk16,Tren16,Spag17,Bila17}
is the most popular playground thanks to its versatility, relative
simplicity, and experimental accessibility with ultracold atoms in
optical potentials
\cite{Lee12,Albi05,Gati07,Este08,Tren16,Spag17}. In this paper we
propose a recipe for generating NOON states \cite{Kok02}, also
known as large cat sates \cite{Cira98}, in the two-site
Bose-Hubbard model. It should be mentioned from the very beginning
that, due to decoherence processes inevitably present in a
laboratory experiment (particle losses, fluctuations of the
optical potential, etc.), the NOON state can be obtained only for
relatively small number of atoms. In other words, in the
thermodynamic limit $N\rightarrow\infty$ one always  gets a
statistical mixture of two states with all atoms localized in
either of two wells -- the phenomenon known as spontaneous
localization or parity-symmetry breaking phase transition. One of
the goals of this work is to estimate the maximal number of atoms
for which one can create the NOON state with the present day
experimental facilities.

Formally, the NOON state is the ground state of the one-dimensional {\em attractive} Bose-Hubbard model \cite{Spek99,Ho04,Lee06} in the strong interaction regime. In practice, however, this state is hard to reach because the NOON state are fragile to particle losses caused  by the collision instability \cite{Dodd96}. To avoid this problem we consider {\em repulsive} atom-atom interactions where the NOON state appears to be the upper energy state of the system. In what follows we show that this state can be reached in the course of adiabatic passage through an excited state quantum phase transition (ESQPT) \cite{Capr08,Pere11,Rela14}. It is generally believed that such an adiabatic passage would require extremely long evolution time,  which scales exponentially with the number of particles \cite{Huan06}. Here, by detailed examination of the system spectrum in a view of the Landau-Zenner tunnelling, we demonstrate that the evolution time actually scales {\em algebraically} with the number of bosons $N$. A pseudo-classical interpretation of the adiabatic passage with the ESQPT corresponding to a separatrix crossing in the classical phase-space is provided.

\section{System overview}
\label{sec1}

\begin{figure}
\includegraphics[width=0.45\textwidth]{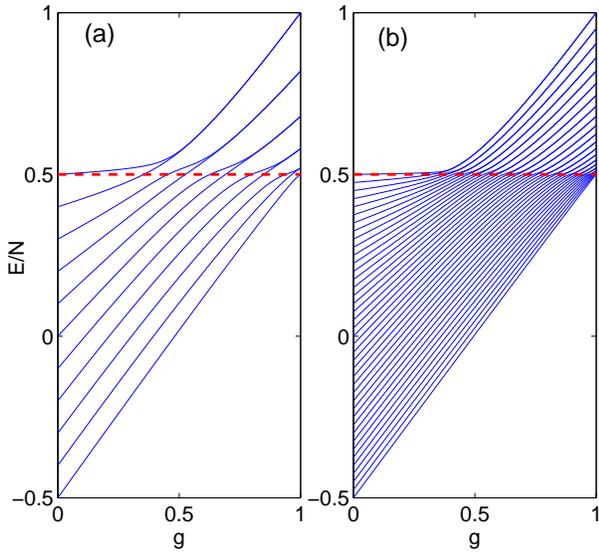}
\caption{Energy spectrum of $N=10$ (left) and $N=40$ (right) bosons against the macroscopic interaction constant $g=UN/2$. (The other parameters are $J=1-g$ and $\delta=0$.) The quantum separatrix  is marked by the red dashed line.}
\label{fig1}
\end{figure}

We consider $N\gg 1$ Bose atoms with repulsive interactions in the
double-well potential. This system is known to be well described
by the two-site  Bose-Hubbard Hamiltonian \cite{Spek99,Java99}
\begin{equation}
\label{a1}
\widehat{H}=-\frac{J}{2} \left( \hat{a}_2^\dag \hat{a}_1 +\hat{a}_2 \hat{a}_1^\dag \right)
+\frac{U}{2}\sum_{l=1,2} \hat{n}_l(\hat{n}_l-1) + \delta(\hat{n}_2-\hat{n}_1) \;,
\end{equation}
where $J$ is the hopping matrix element, $U$ the microscopic interaction constant, $\hat{a}_l$ and $\hat{a}_l^\dagger$ the bosonic annihilation and creation operators,  $\hat{n}_l$ the number operator, and $\delta$ the difference between the on-site energies. For $N$ bosons the Hilbert space of the Hamiltonian (\ref{a1}) of dimension ${\cal N}=N+1$ is spanned by Fock states
\begin{equation}
\label{a2}
|N_1,N_2 \rangle=|N/2-n,N/2+n \rangle \equiv | n \rangle \;,\quad  |n|\le N/2 \;,
\end{equation}
where $N_1 + N_2=N$. Above we used a symmetric parameterization to label the Fock  states by a single quantum number $n$ ($N$ is assumed to be even). The full spectrum of the system is shown in Fig.\ref{fig1}, where we introduced  the macroscopic interaction constant $g=UN/2$ and set the hopping matrix element to $J=1-g$. Thus, the case $g=0$ corresponds to the system of non-interacting bosons while in the case $g=1$ the inter-well tunneling is completely suppressed. It is easy to prove that the spectrum is equidistant for $g=0$ and quadratic for $g=1$, with all energy levels except the ground state being twofold degenerate \cite{Muel06}. The spectrum for intermediate $g$ possesses the quantum separatrix and can be understood by employing the pseudo-classical approach which we review in Sec.~\ref{sec2}.

\begin{figure}[t]
\includegraphics[width=0.45\textwidth]{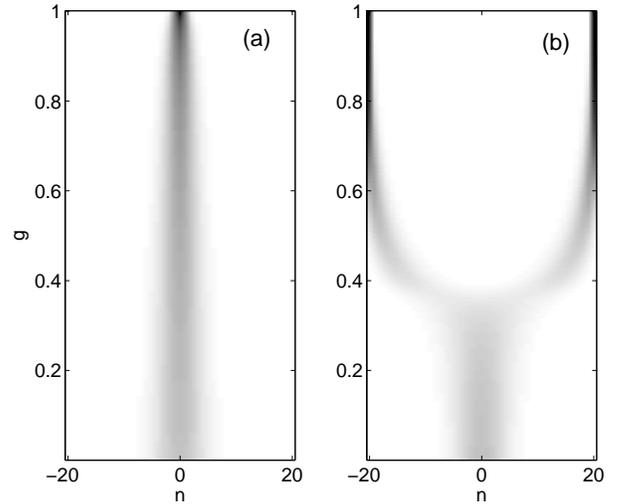}
\caption{Squared absolute values of expansion coefficients
Eq.~(\ref{a5}) of the ground (left) and upper energy (right)
states against the macroscopic interaction constant $g$.}
\label{fig2}
\end{figure}

Among the eigenstates $|\Psi_j\rangle$ of the Hamiltonian (\ref{a1}) of particular interest are the states with minimal and maximal energy.  For $g=0$ the ground state of the system is a Bose-Einstein condensate with all particles occupying  the symmetric single-particles state,
\begin{equation}
\label{a3}
|\Psi_0(g=0)\rangle=\frac{1}{\sqrt{2^NN!}}\left(a_1^\dag+a_2^\dag\right)^N|{ vac}\rangle \;,
\end{equation}
while the upper energy state is a Bose-Einstein condensate with all particles occupying  the antisymmetric single-particles state,
\begin{equation}
\label{a4}
|\Psi_N(g=0)\rangle=\frac{1}{\sqrt{2^NN!}}\left(a_1^\dag-a_2^\dag\right)^N|{vac}\rangle \;.
\end{equation}
Let us follow these states under variation of $g$. At each value of $g$ eigenfunctions are found as an expansion over the Fock states (\ref{a2}),
\begin{equation}
\label{a5}
|\Psi_j(g)\rangle = \sum_{n=-N/2}^{N/2} c^{(j)}_n(g) | n \rangle \;,
\end{equation}
For $j=0$ and $j=N$ the results are shown in Fig \ref{fig2}. It is
seen that the ground state transforms into the fragmented
condensate \cite{Muel06}
\begin{equation}
\label{a6}
|\Psi_0(g=1) \rangle = |N/2,N/2\rangle  \;,
\end{equation}
while the upper energy state evolves into the NOON state
\begin{equation}
\label{a7}
|\Psi_N (g=1) \rangle = |NOON\rangle \equiv \frac{1}{\sqrt{2}}\left( |N,0\rangle + |0,N\rangle \right)  \;.
\end{equation}

%
\begin{figure}[t]
\includegraphics[width=0.5\textwidth]{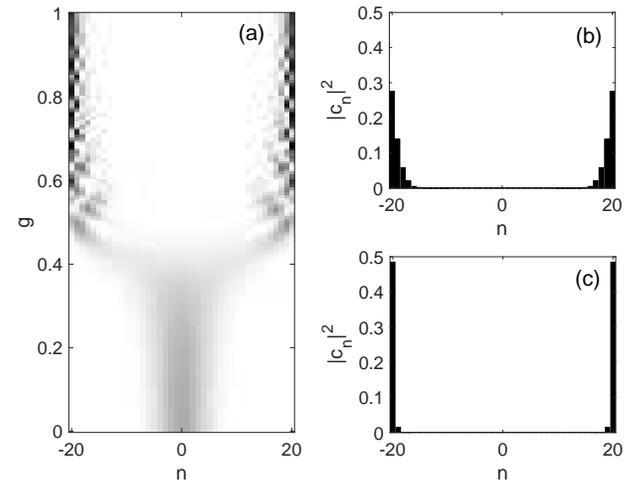}
\caption{Panel (a): Squared absolute values of expansion
coefficients over the Fock basis as the function of $g=\nu t$ for
the adiabatic passage with $\nu=0.1$. Panels (b) and (c) compares
final state of the system for $\nu=0.1$ and $\nu=0.025$.}
\label{fig3}
\end{figure}

Next we consider time evolution of the system according to the Schr\"odinger equation,
\begin{equation}
\label{a8}
i\frac{{\rm d}}{{\rm d} t} |\psi \rangle= \widehat{H}(g) |\psi \rangle  \;,\quad J=1-g \;,
\end{equation}
with the interaction constant $g$ growing linearly from $0$ to $1$
during the time interval $T=1/\nu$. In Fig.~\ref{fig3}(a) we
present the results of numerical simulations of the system
dynamics for $|\psi(t=0) \rangle= | \Psi_N(g=0) \rangle$ and
$\nu=0.1$.
 Shown are squared absolute values of the expansion coefficients $c_n(t)=\langle  n |\psi(t)\rangle$.
 One can see in Fig.~\ref{fig3}(b) that the final state $|\psi(t=T)\rangle$ does not ideally coincide  with the target NOON state Eq.~(\ref{a7}).
 With a smaller ramp rate, however, the result is almost perfect,  see Fig.~\ref{fig3}(c). In the next sections we analyze the discussed adiabatic passage in more detail and quantify  the final state  $| \psi(t=T) \rangle$.
 To pay credits to other works we mention that adiabatic passage for the ground state of the attractive Bose-Hubbard model was considered earlier
 in Ref.~\cite{Lee06,Lee09} and a different adiabatic passage, which involves the rising potential barrier which separates a  Bose-Einstein condensate
 into two parts, was analyzed in Ref.~\cite{Java99,Meno01,Pezz05,Stre07}.

\section{Pseudo-classical approach}
\label{sec2}

To get a deeper insight into the discussed adiabatic passage we resort to the pseudo-classical approach. This approach borrows its ideas from the semi-classical method in single-particle quantum mechanics to address the spectral and dynamical properties of the system of $N$ interacting bosons, with $1/N$ playing the role of Planck's constant \cite{Mahm05,Moss06,Grae07,Zibo10}. Formally, the creation and annihilation operators are substituted with $C$-numbers as $\hat{a}_l/\sqrt{N} \rightarrow a_l$ and $\hat{a}_l^\dagger/\sqrt{N}\rightarrow a_l^*$, which also implies rescaling of the Hamiltonian as $\widehat{H}/N \rightarrow H$. For the two-site Bose-Hubbard model this leads to the classical Hamiltonian \cite{Smer97}
\begin{equation}
\label{b1}
H= \frac{g}{2} I^2  - \frac{J}{2} \sqrt{1-I^2}\cos\phi \;, \quad J=1-g \;,
\end{equation}
where $g=UN/2$ is the macroscopic interaction constant.

\begin{figure}[b]
\includegraphics[width=0.5\textwidth]{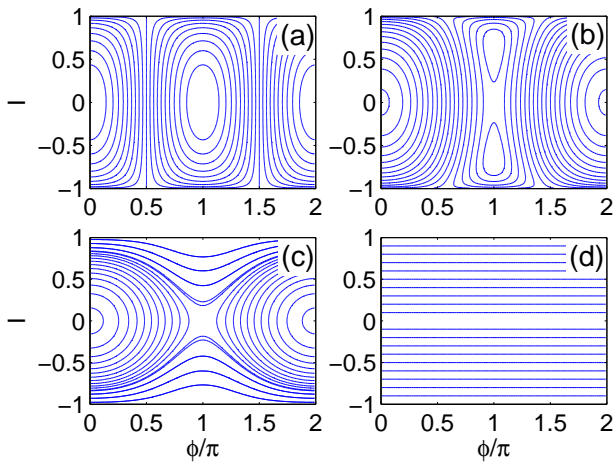}
\caption{Phase portraits of the classical Hamiltonian (\ref{b1})
for (a) $g=0$, (b) $g=0.4$, (c) $g=0.7$, and (d) $g=1$.}
\label{fig4}
\end{figure}

Fig.~\ref{fig4} shows the phase portrait of the system (\ref{b1})
at four different values of $g$:  $0$, $0.4$, $0.7$, and $1$. By
using the relation $I=\sin \theta$ phase portraits of the system
can be also drawn on sphere of the unit radius. In this
representation the line $I=1$ ($I=-1$) reduces to single  point --
the north (south) pole of the sphere. For $g=0$ the phase portrait
contains two elliptic points at $(I,\phi)=(0,0)$ (minimal energy)
and  $(I,\phi)=(0,\pi)$ (maximal energy), see Fig.\ref{fig4}(a).
As $g$ is increased above $g_{cr}=1/3$ the latter elliptic point
bifurcates into two elliptic points at  $(I,\phi)=(\pm I^*,\pi)$,
where $I^*$ is a function of $g$. With a further increase of $g$
the island around the point $(I,\phi)=(0,0)$ vanishes while the
islands around $(I,\phi)=(\pm I^*,\pi)$ monotonically grow,
finally leading to the phase-space portrait shown in
Fig.\ref{fig4}(d).

The depicted phase portraits suffice to find the energy spectrum
shown in Fig.~\ref{fig1} by using the semiclassical quantization
rule, where the phase volume encircled by a trajectory is required
to be a multiple of the effective Planck constant $h=1/N$. Then
the central island around the point $(I,\phi)= (0,0)$ gives energy
levels below the quantum separatrix while two symmetric islands
around $(I,\phi)= (\pm I^*,\pi)$ give degenerate levels above the
quantum separatrix. The details  are given in Ref.~\cite{Grae07},
where it was demonstrated that the pseudo-classical approach
provides an accurate approximation to the exact spectrum even for
$N=10$.

\begin{figure}
\includegraphics[width=0.5\textwidth]{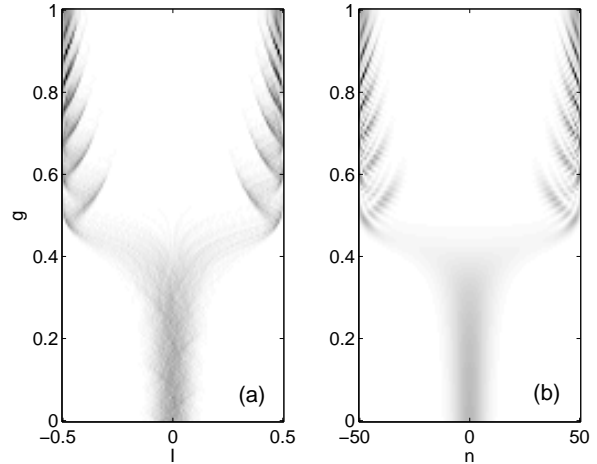}
\caption{Comparison between the classical (left panel) and quantum
(right panel) dynamics. Parameters are $\delta=0$, $\nu=0.1$, and
$N=100$.} \label{fig5}
\end{figure}

Let us now study dynamics of the classical system (\ref{b1}) when
both $g$ and $J$ vary in time as $g=\nu t$ and $J=1-g$. As the
initial condition we take an ensemble of particles with the
probability distribution given by the two-dimensional Gaussian
centered at the elliptic point $(I,\phi)=(0,\pi)$. For comparison
with quantum dynamics the width of the Gaussian is adjusted to
$\sigma=\sqrt{N}$.  The left panel in Fig.~\ref{fig5} shows the
evolution of the classical distribution function $\rho(I,t)$ for
$N=100$. (We stress one more time that the latter parameter
determines only the width of the initial distribution.) The left
panel in Fig.~\ref{fig5} should be compared with the right panel
showing the quantum evolution. The observed agreement underlines
the classical phenomenon behind the quantum results discussed in
the previous section. Classically, the particles are captured into
the upper and lower islands emerging after bifurcation of the
elliptic point $(I,\phi)=(0,\pi)$ and then transported towards
$I=1$ and $I=-1$, respectively. The phenomenon of capturing into
(and releasing from) an elliptic island was considered earlier in
Ref.~\cite{32} in a different context. It involves the crossing of
instantaneous separatrix that, in turn, was analyzed in
Ref.~\cite{Hann86}.

\begin{figure}
\includegraphics[width=0.45\textwidth]{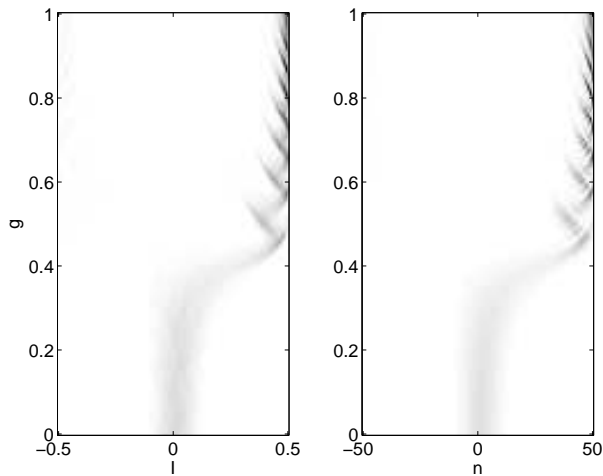}
\caption{Same as in Fig.~\ref{fig5} with $\delta=0.01$.}
\label{fig6}
\end{figure}

To conclude this section we discuss the effect of non-zero $\delta$. For $\delta\ne0$ the emerging islands have different size, which makes  $\rho(I,t)$ asymmetric with respect to $I\rightarrow -I$. To characterize this asymmetry we introduce the population imbalance
\begin{equation}
\label{b4a}
G = \int_{0}^{1} \rho(I,T){\rm d}I -  \int_{-1}^{0} \rho(I,T){\rm d}I   \;.
\end{equation}
If $\delta$ is increased the population imbalance (\ref{b4a}) grows  monotonically, approaching $|G|=1$, see Fig.~\ref{fig6}. Importantly, the imbalance also grows if $\nu$ is decreased and for any finite $\delta$ the imbalance is unity in the limit $\nu\rightarrow 0$.

\section{Landau-Zener tunneling and fidelity to the NOON state}
\label{sec3}

\begin{figure}[t]
\includegraphics[width=0.5\textwidth]{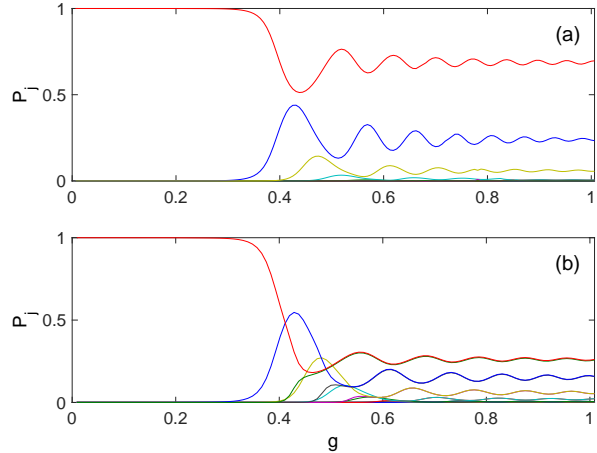}
\caption{Populations of the instantaneous energy levels  for $N=30$,  $\nu=0.1$, and $\delta=0$ (upper panel) and $\delta=0.0001$ (lower panel).}
\label{fig7}
\end{figure}

In the previous section we explained the quantum results depicted in Fig.~\ref{fig2} by using the pseudo-classical approach. The quantum-mechanical explanation of these results is based on the notion of Landau-Zenner tunneling. Due to this phenomenon several energy levels become populated as we follow the upper most level in Fig.~\ref{fig1} with a finite sweeping rate. This is illustrated in Fig.~\ref{fig7}(a) which shows the populations of the instantaneous energy levels $P_j(t)$,
\begin{equation}
\label{c1}
P_j(t)=|\langle \Psi_j(g=\nu t) | \psi(t) \rangle|^2     \;,
\end{equation}
for $\nu=0.1$ and $N=30$. Notice that only even levels are populated because of different symmetry of eigenstates  of the Hamiltonian (\ref{a1}) with odd and even index $j$.  To quantify the effect of Landau-Zener tunneling we introduce the fidelity
\begin{equation}
\label{b4}
F = |\langle NOON|\psi(T)\rangle|^2     \;,
\end{equation}
which characterizes how close the final state is to the target NOON state.  In the limit $\nu\rightarrow0$ it is enough to take into account only the nearest high energy level of the same (even) symmetry, which alone determines fidelity of the final state through the celebrated Landau-Zener equation
\begin{equation}
\label{b5}
F= 1-\exp\left(-\frac{\pi \Delta^2}{2|\alpha| \nu}\right)    \;.
\end{equation}
In Eq.~(\ref{b5}) $\Delta$ is the energy gap between the upper
most level and the next level of the same symmetry, $\nu=1/T$ the
sweeping rate, and  $\alpha$ is determined by the angle at which
two levels approach each other. Since the energy gap $\Delta$ and
$|\alpha|$ scales algebraically with $1/N$, we expect that the
evolution time $T$ has to be increased proportionally to the
number of particles to insure a given fidelity. Direct numerical
simulations of the adiabatic passage for different $N$ confirm
this hypothesis, see Fig.~\ref{fig8}(a). It is interesting to
discuss the depicted result with respect to the recent laboratory
experiment \cite{Tren16} which studies the
parity-symmetry-breaking phase transition for $N\approx4500$
attractively interacting atoms in a double-well potential. Taking
$J/h=40Hz$ and the evolution time $\sim1s$ we get $N\approx 30$
and this number can be easily increased by relaxing the fidelity
to $F=0.9$ and using a time-dependent sweeping rate $\nu=\nu(t)$
that optimizes the adiabatic passage. We stress that the above
estimate is obtained under the assumption of negligible
decoherence processes which we shall discuss in Sec.~\ref{sec4}.

Next we analyze the effect of non-zero $\delta\ll J$ in the
Hamiltonian (\ref{a1}) from the quantum-mechanical viewpoint.
Non-zero $\delta$ breaks the reflection symmetry of the system, so
that eigenstates of the Hamiltonian (\ref{a1}) at $g\gg J$ are
given by the Fock states $| N/2-n,N/2+n \rangle$ and   $|
N/2+n,N/2-n \rangle$ but not their symmetric or antisymmetric
superpositions. (In particular, $|\Psi_N \rangle \approx
|N,0\rangle$ and $\Psi_{N-1} \rangle \approx |0,N\rangle$.) This
drastically changes Fig.~\ref{fig7}(a) -- now both odd and even
instantaneous energy levels become populated during the adiabatic
passage, see Fig.~\ref{fig7}(b). For the considered extremely
small value of $\delta$ this difference simply reflects a change
of the basis and, physically, both Fig.~\ref{fig7}(a) and
Fig.~\ref{fig7}(b) describe the same process, which results in the
NOON state as the final state of the system. However, for a larger
$\delta$ we see considerable deviation from the NOON state, see
Fig.~\ref{fig8}(b).  In particular, in full analogy with the
classical result, the population imbalance $|G|$ approaches the
unity if $|\delta|$ is increased.

\section{Decoherence effects}
\label{sec4}

\begin{figure}[t]
\includegraphics[width=0.5\textwidth]{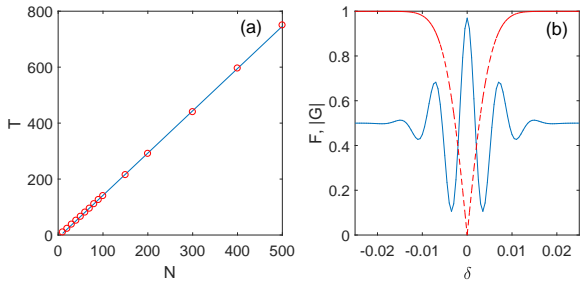}
\caption{Left panel:  Minimal evolution time $T$ insuring fidelity $F=0.99$ versus the number of bosons $N$. Right panel: Population imbalance $|G|$ (dashed line) and fidelity $F$ (solid line) as the function of $\delta$ for $N=40$ and $\nu=0.025$.}
\label{fig8}
\end{figure}

The result depicted  in Fig.~\ref{fig8}(a) proves that, at least in principle, one can create arbitrary large cat state by simply increasing the duration of the adiabatic passage proportionally to the number of particles $N$. This, however, implicitly assumes the absence of any decoherence process \cite{Huan06} and precision control over the system parameters, in the first place, over parameter $\delta$.  In this section we discuss decoherence  caused by fluctuations of $\delta$, which are unavoidable in a laboratory experiment.

\begin{figure}
\includegraphics[width=0.4\textwidth]{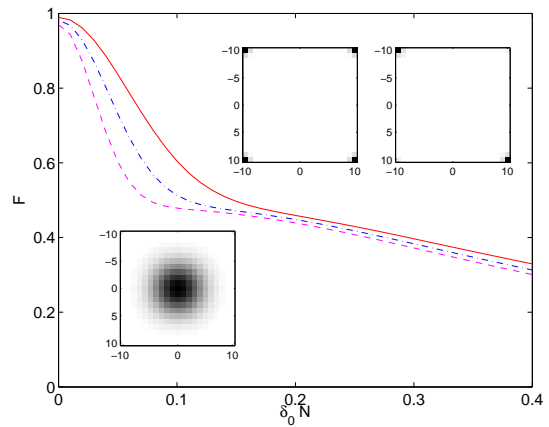}
\caption{Fidelity (\ref{b7}) as the function of the noise amplitude  for $N=10$ and $\nu=0.1$ (solid line), $N=20$ and $\nu=0.05$ (dash-dotted line), and $N=40$ and $\nu=0.025$ (dashed line). Insets show the initial density matrix for $N=20$ (lower-left conner) and final density matrices for $\delta_0=0$ and $\delta_0=0.1/N$ (upper-right conner).}
\label{fig9}
\end{figure}

In the presence of fluctuations the fidelity(\ref{b4}) should be redefined as
\begin{equation}
\label{b7}
F = \langle NOON |{\cal R}(T)| NOON\rangle    \;,\quad
{\cal R}(t)=\overline{ |\psi(t)\rangle\langle \psi(t) | } \;,
\end{equation}
where the bar denotes the average over fluctuations. To be specific, we assume that $\delta(t)$ is the white noise with vanishing mean value, i.e., $\overline{\delta(t)\delta(t')}=\delta_0^2\delta(t-t')$. Then the density matrix ${\cal R}(t)$  is easy to show to obey the following master equation \cite{28}
\begin{equation}
\label{b8}
\frac{{\rm d} {\cal R}}{{\rm d} t}= -i[\widehat{H},{\cal R}]
-\delta_0^2\left[\hat{n},[\hat{n},{\cal R}]\right] \;,
\end{equation}
where $\hat{n}=\hat{n}_1-\hat{n}_2$. We solve Eq.~(\ref{b8}) for the adiabatic passage discussed above. Fig.~\ref{fig9} shows  fidelity (\ref{b7}) as the function of the noise amplitude $\delta_0$ for three system sizes $N=10,20,40$, where we proportionally decreased the sweeping rate $\nu$ to insure fidelity $F\approx 1$.  One striking feature of the depicted functions is a rapid decay of fidelity to $F\approx 0.5$ in the  interval  $0<\delta_0<\delta_0^*$ where $\delta_0^* = \delta_0^*(N)$. In this interval the off-diagonal elements of the density matrix ${\cal R}(T)$ gradually vanish. On the other hand, the diagonal elements of the density matrix remain essentially unaffected. Clearly, this result illustrates the usual quantum-to-classical transition due to a decoherence process \cite{Ditt88,Zure91,28}. Notice that the larger system is, the more it is sensitive to decoherence. Numerical results depicted in Fig.~\ref{fig9} indicate that $\delta_0^*$ decreases with $N$ faster than $1/N$.

Next we briefly discuss decoherence due to particle losses. In the
case of not conserved number of particles the master equation for
the system density matrix reads
\begin{equation}
\label{b9}
\frac{{\rm d} {\cal R}}{{\rm d} t}= -i[\widehat{H},{\cal R}]
-\gamma \sum_{l=1,2}
(\hat{a}^\dagger_l\hat{a}_l{\cal R}-2\hat{a}_l{\cal R}\hat{a}^\dagger_l + {\cal R}\hat{a}^\dagger_l\hat{a}_l) \;,
\end{equation}
where $\gamma$ denotes the decay rate (see, for example,
Ref.~\cite{Kord15}). The value of $\gamma$ in Eq.~(\ref{b9})
crucially depends on the sign of interatomic interactions. For
example, in the already sited experiment \cite{Tren16} with
attractively interacting atoms the decay rate was $\gamma\approx
0.1$ which makes impossible generation of the NOON state even for
$N\sim 10$ atoms. On the other hand, it is known that a
Bose-Einstein condensate of repulsively interacting atoms may have
the life-time up to few hours that assumes $\gamma\sim 10^{-4}$
\cite{Corn02}. The negligible decoherence rate due to particle
losses is our main reason for considering the adiabatic passage
for the upper energy state of repulsively interacting atoms
instead of that for the ground state of attractively interacting
atoms. In all other aspects there is no conceptional difference
between the adiabatic passages for the ground and upper states.

\section{Preparation of the excited state}
\label{sec5}

\begin{figure}[t]
\centering
\includegraphics[width=0.4\textwidth]{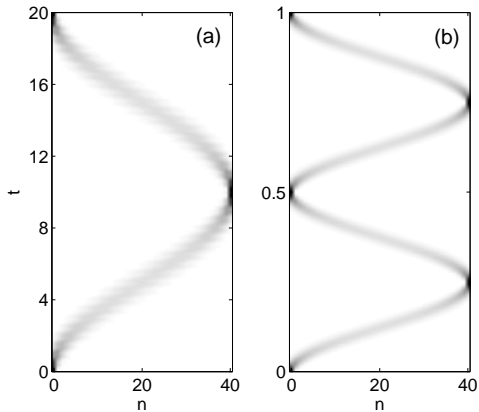}
\caption{Populations of eigenstates of the Hamiltonian (\ref{a1}) as the function of time. (Note that for $g=0$ the eigenstates of (\ref{a1}) are given by $|N-n,n\rangle$ where $n$ now denotes the number of particles in the antisymmetric single-particle state.) Parameters are $N=40$, $g=0$, $J=1$, $\omega=J$, $\delta_0=0.05$ (left panel) and $N=40$, $g=0$, $J=0.01$, $\delta=1$ (right panel). \vspace{0.4cm}}
\label{fig10}
\end{figure}

Finally, we discuss a method to excite the system of non-interacting bosons ($g=0$) into the highest energy state. A way to do this is to drive the system by periodically changing parameter $\delta$ as $\delta(t)=\delta_0 \sin( \omega t)$, where the frequency $\omega$ coincides with the transition frequency between the symmetric and antisymmetric single-particle states uniquely determined by the parameter $J$. If $\delta_0 \ll J$ (the latter condition justifies the rotating-wave approximation) the problem can be solved analytically and leads to the Rabi oscillations, see Fig.~\ref{fig10}(a). Thus, to excite the system in the upper state, we need to drive it for one half of the Rabi period.

Another, perhaps even simpler way to obtain the excited state (\ref{a4}) is to quench the system into the parameter region $\delta\gg J$ by suddenly tilting the double-well. Then the time evolution of the expansion coefficients is approximately given by $c_n^{(j)}(t)=\exp(i2\delta n t) c_n^{(j)}(0)$ and after one half of the period $T_B=\pi/\delta$ (which can be interpreted as the Bloch period) the state (\ref{a3}) transforms into the state (\ref{a4}), see Fig.~\ref{fig10}(b).

\section{Conclusions}
\label{sec6}

We suggested a method for creating the NOON state of Bose atoms,
i.e., coherent  superposition of two states in which all particles
are in the same well of the double-well potential.  Unlike to
previous studies, which almost exclusively focused on the case of
attractive interactions \cite{remark}, we considered the
repulsively interacting atoms that avoids the problem of particle
losses. The scheme protocol consists of two steps. First, by
setting the inter-atomic  interactions to zero we transfer the
system from the ground state to the upper excited state. Second,
adiabatically increasing the interaction strength and
simultaneously decreasing the hopping rate we  transform this
excited state to the NOON state. In the Fock space the latter
stage can be viewed as splitting of the initially localized wave
packet into two packets \cite{Stre09}. This process was shown to
have a pseudo-classical counterpart and some of quantum results,
for example,  the population  imbalance $G$ can be obtained by
using pure classical arguments. Of course, the classical approach
cannot address phase coherence between the packets, which is
characterized by the fidelity $F$.

Formally, the suggested scheme allows us to create an arbitrary large cat state. However, any experimental  realization of the scheme protocol imposes fundamental limitation on the number of atoms due to decoherence processes present in a laboratory experiment. Here, we analyzed the decoherence caused  by fluctuation of the parameter $\delta$ (the energy mismatch between the left and right wells of the double-well potential) that appears to be crucial  for the system dynamics.  It was shown that there is a critical value for the fluctuation amplitude $\delta_0^*\sim 1/N$ above which the final state of the system becomes  `classical NOON state', i.e., incoherent superposition of two states in which all particles are in the same well of the double-well potential.  Thus to get the NOON state with a large number of atoms every effort to reduce fluctuation of $\delta$ should be taken.

{\em Acknowledgements.}
The authors acknowledge financial support from Russian Foundation for Basic Research, Government
of Krasnoyarsk Territory, and Krasnoyarsk Region Science and Technology Support Fund through the grant No. 16-42-240746.


\begin{thebibliography}{99}
\bibitem{remark}
Here we mention the other methods for creating the NOON state of
repulsively interacting atoms based on the quench \cite{Stre09} or
lossy dynamics \cite{Kord12}, and the works
\cite{Mich03,Pezz09,Gros12,Ried12,Strob14} which explore
possibility of using the two-component BEC in a single well/trap
instead of the single component BEC in a double-well potential.
\end{thebibliography}


\end{document}